# QSDsan: An Integrated Platform for Quantitative Sustainable Design of Sanitation and Resource Recovery Systems

Yalin Li[a,b†*], Xinyi Zhang[c†], Victoria L. Morgan[a], Hannah A.C. Lohman[c], Lewis S. Rowles[a,1], Smiti Mittal[d], Anna Kogler[e], Roland D. Cusick[c], William A. Tarpeh[e,f], Jeremy S. Guest[a,b,c]

[a] Institute for Sustainability, Energy, and Environment, University of Illinois Urbana-Champaign, 1101 W. Peabody Drive, Urbana, IL 61801, USA.

[b] DOE Center for Advanced Bioenergy and Bioproducts Innovation, University of Illinois Urbana-Champaign, 1206 W. Gregory Drive, Urbana, IL 61801, USA.

[c] Department of Civil and Environmental Engineering, University of Illinois Urbana-Champaign, 3221 Newmark Civil Engineering Laboratory, 205 N. Mathews Avenue, Urbana, IL 61801, USA.

[d] Department of Bioengineering, Stanford University, 129 Shriram Center, 443 Via Ortega, Stanford, California 94305, USA.

[e] Department of Civil and Environmental Engineering, Stanford University, 311 Y2E2, 473 Via Ortega, Stanford, California 94305, USA.

[f] Department of Chemical Engineering, Stanford University, 129 Shriram Center, 443 Via Ortega, Stanford, California 94305, USA.

[1] Present address: Civil Engineering and Construction, Georgia Southern University, 201 COBA Drive, BLDG 232 Statesboro, GA 30458, USA.

**Email addresses:**

yalinli2@illinois.edu (Y.L.); xinyiz6@illinois.edu (X.Z.);

vlmorgan@illinois.edu (V.L.M.); hlohman2@illinois.edu (H.A.C.L.);

lrowles@georgiasouthern.edu (L.S.R.); smiti06@stanford.edu (S.M.);

akogler@stanford.edu (A.K.); rcusick@illinois.edu (R.D.C.);

wtarpeh@stanford.edu (W.A.T.); jsguest@illinois.edu (J.S.G)

**†Co-first authors:** Y. Li and X. Zhang contributed equally to this work.

**Corresponding author:** yalinli2@illinois.edu, +1 (217) 418-4672

**Graphical abstract**

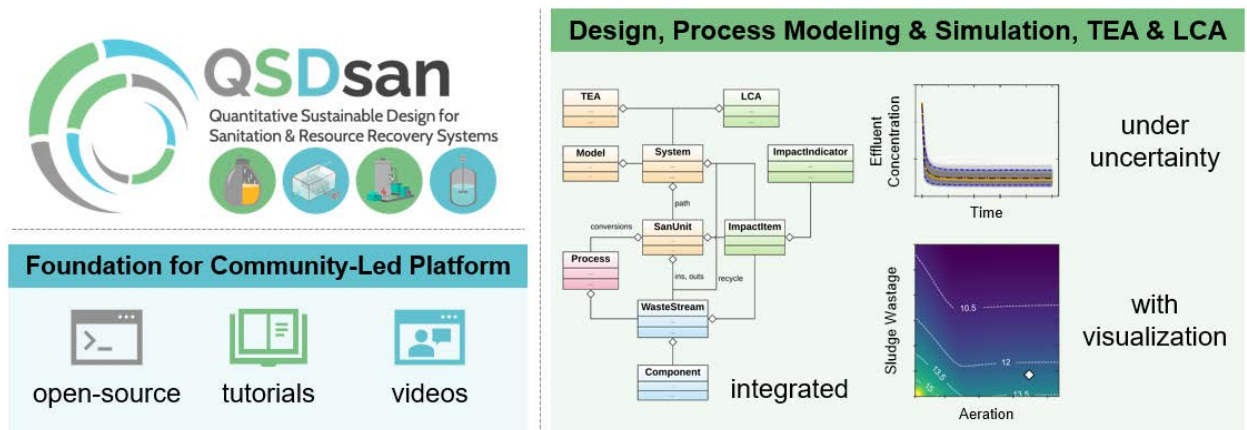

**Highlights**

- QSDsan is an open-source platform for sanitation and resource recovery systems
- QSDsan integrates system design, simulation, and sustainability characterization
- We show QSDsan's capacity via TEA, LCA, and dynamic simulation under uncertainty
- We support QSDsan with online documentation, tutorials, and transparent management

**Abstract**

Sustainable sanitation and resource recovery technologies are needed to address rapid environmental (e.g., climate change) and socioeconomic changes (e.g., population growth, urbanization). Research prioritization is critical to expedite the development and deployment of such technologies across their vast system space (e.g., technology choices, design and operating decisions). In this study, we introduce QSDsan – an open-source tool written in Python (under the object-oriented programming paradigm) and developed for the quantitative sustainable design (QSD) of sanitation and resource recovery systems. As an integrated platform for system design, process modeling and simulation, techno-economic analysis (TEA), and life cycle assessment (LCA), QSDsan can be used to enumerate and investigate the opportunity space for emerging technologies under uncertainty, while considering contextual parameters that are critical to technology deployment. We illustrate the core capabilities of QSDsan through two distinct examples: (i) evaluation of a complete sanitation value chain that compares three alternative systems; and (ii) dynamic simulation of the wastewater treatment plant described in the benchmark simulation model no. 1 (BSM1). Through these examples, we show the utility of QSDsan to automate design, enable flexible process modeling, achieve rapid and reproducible simulations, and to perform advanced statistical analyses with integrated visualization. We strive to make QSDsan a community-led platform with online documentation, tutorials (explanatory notes, executable scripts, and video demonstrations), and a growing ecosystem of supporting packages (e.g., DMsan for decision-making). This platform can be freely accessed, used, and expanded by researchers, practitioners, and the public alike, ultimately contributing to the advancement of safe and affordable sanitation technologies around the globe.

## 1. Introduction

With the increasing pace of technology development[1–3] and growing complexity of sustainability challenges,[4–7] there is a need for robust and agile tools to quickly identify critical barriers, prioritize research opportunities, and navigate multi-dimensional sustainability tradeoffs in the research, development, and deployment (RD&D) of technologies.[8–10] This need is particularly pressing for the field of sanitation as it concerns one of the most basic human rights, which directly addresses the sixth Sustainable Development Goal (SDG)[11] proposed by the United Nations (universal sanitation by 2030), and is connected to many other SDGs (e.g., resource circularity, carbon neutrality). To improve sanitation service coverage in resource-limited communities, a portfolio of technologies are needed, including those that do not require large capital investment (e.g., via non-sewered sanitation) and that lower costs and environmental impacts through resource recovery.[12,13]

Toward this end, multiple high-fidelity commercial software are available for the design and simulation of water and wastewater systems as well as resource recovery technologies (e.g., GPS-X™,[14] SUMO©,[15] BioWin,[16] WEST,[17] Aspen Plus®,[18] SuperPro Designer[19]). However, these tools primarily focus on conventional, centralized technologies rather than early-stage RD&D of novel, decentralized systems. Moreover, the current approach of segregating system design, simulation, and sustainability characterization (e.g., techno-economic analysis, TEA; life cycle assessment, LCA) into multiple tools (e.g., GPS-X™ for design and simulation; CapdetWorks[20] for TEA and SimaPro[21] for LCA) creates challenges in execution and maintaining transparency, motivating the development of new tools to streamline this workflow.[9] Further, the lack of support for uncertainty and sensitivity analyses often limits the scope of existing studies to a narrow set of design or control decisions and select combinations of technological and contextual parameters, thus undermining the utility of these tools for early-stage technologies. Although features have been included in some software to enable incorporation of uncertainty to a limited degree (e.g., by allowing advanced simulation settings in programming languages like C# or Python), it remains challenging to perform robust uncertainty and sensitivity analyses with these tools because procedures beyond batch simulation (e.g., parameter sampling, calculation of sensitivity indices, statistical analysis, visualization) still need to be executed externally.

To address these gaps, we herein present QSDsan – an open-source tool that leverages the quantitative sustainable design (QSD) methodology for integrated design, simulation, and sustainability evaluation of sanitation and resource recovery systems.[10] Built under the object-oriented programming (OOP) paradigm using Python (3.8+), QSDsan aims to support and expedite RD&D of early-stage sanitation technologies as a community-led platform that provides flexible, transparent, and freely accessible modules for modeling and evaluation. With a rich collection of Python libraries and the embracement of open source by the rapidly growing scientific programming community, QSDsan has the potential to continuously advance together with emerging sanitation and resource recovery technologies.

The main features of QSDsan include bulk property calculations of waste streams, equilibrium and dynamic process modeling, user-defined unit operation design, automated system simulation, integrated TEA and LCA, and advanced uncertainty and sensitivity analyses with built-in visualization functions. In addition to introducing the underlying structure of QSDsan, we highlight its capabilities through two illustrative implementations under different (equilibrium vs. dynamic) simulation modes. In the first implementation (equilibrium mode), we simulated three alternative sanitation systems under uncertainty and characterized their sustainability via TEA and LCA, where each alternative includes human excreta input, user interface and onsite storage, conveyance, centralized treatment, and reuse of treated and recovered excreta-derived products.[25,26] In the second implementation (dynamic mode), we evaluated the system described in benchmark simulation model no.1 (BSM1),[27,28] which consists of a five-compartment activated sludge reactor (two anoxic tanks and three aerobic tanks, all modeled with the activated sludge model no. 1, ASM1[29]) and a secondary clarifier (modeled as a 10-layer non-reactive unit).[30] Finally, we discuss how QSDsan can be continuously developed (e.g., by connecting with other tools for decision-making and system optimization) to better contribute to the advancement of sustainable sanitation and resource recovery systems.

## 2. Methods

### *2.1. Structure of QSDsan*

To enable the modeling of any sanitation technologies and systems, QSDsan leverages the structure proposed in BioSTEAM[31,32] – an agile platform developed for the design, simulation, and TEA of biorefineries. In Python, the OOP paradigm has two core concepts – "classes" and "instances", both of which can be referred to as "objects". Different classes can be established to provide pre-defined sets of data and/or methods (i.e., functional algorithms), which are collectively referred to as "attributes". For each class, subclasses can be created to inherit or modify its attributes (e.g., a lagoon class can have anaerobic and facultative subclasses). In application, "instances" of classes are created in the systems established by the user, where these "instances" are the actual implementation of the classes with inherited attributes. For example, users can design a system with any number of anaerobic lagoons, and each of these lagoons will be an instance of the anaerobic lagoon class. By inheriting from BioSTEAM[31,32] and Thermosteam (BioSTEAM's thermodynamic engine)[33,34] under this OOP paradigm, QSDsan is capable of rigorous and automated process modeling with enhanced features geared toward sanitation and resource recovery applications (**Figure 1**).

#### 2.1.1. Tracking mass and energy flows

When using QSDsan, users start with creating *Component* (italic words denote modules, classes, or attributes in Python hereinafter) objects, which contain attributes (e.g., nitrogen content, total suspended solids, chemical oxygen demand [COD]-to-mass ratio) relevant to wastewater treatment. *Component* objects can be linked to pure chemicals (e.g., acetic acid) in Thermosteam's[33] database to enable thermodynamic property calculation and simulation (e.g., phase transition). Alternatively, users can also create *Component* objects by selecting from the built-in set of typical components in wastewater treatment modeling and tailoring them to their needs.[35]

With *Component* objects, *WasteStream* objects can be created to handle mass and energy flows (by tracking the quantity, phase, temperature, and pressure of individual components) as well as to calculate bulk properties (e.g., the concentration of volatile suspended solids). A *WasteStream* object can be created by either (i) defining the flowrates of individual *Component* objects or (ii) through established influent characterization models[36] (e.g., based on total COD and COD fractions) with the default component set. Kinetic interactions among components are captured using *Process* objects, which store data on stoichiometries and rate equations. The *Process* class is equipped with algorithms for automatic calculations of unknown stoichiometric coefficients based on conservation of materials (e.g., carbon, nitrogen, COD, charge). With multiple *Process* objects, kinetic models (e.g., ASM1[29]) can be represented as ordinary differential equations (ODEs) describing the rate of production or consumption for each component.

## 2.1.2. Design and simulation of unit operations and systems

The *SanUnit* class is used to design and simulate unit operations (e.g., a bioreactor). Influents and effluents of a *SanUnit* object are represented by *WasteStream* objects. Transformation of *Component* from influent to effluent *WasteStream* can be modeled in either equilibrium (by defining conversions) or dynamic (through *Process* objects) mode. Design (e.g., reactor height, volume), cost (e.g., capital, operational), and utility usage (e.g., heating, cooling) of *SanUnit* objects are stored as attributes of *SanUnit*. These attributes can be fixed at certain values or calculated with respective algorithms using *WasteStream* (e.g., calculate reactor volume based on flowrate and retention time) and *Process* (e.g., calculate electricity usage based on the air flowrate modeled by the aeration *Process*) inputs.

*SanUnit* objects can be connected by *WasteStream* objects and aggregated into *System* objects. In equilibrium mode, mass and energy of influent and effluent *WasteStream* of all *SanUnit* within the *System* are converged at the given conditions. In the dynamic mode, ODEs representing the accumulation rate of components in each *SanUnit* are compiled into *System*-wise ODEs and integrated from the initial conditions over the desired period. After convergence of the *System*, design algorithms provided by the user are simulated to update the system inventory (including chemical and material usage as well as emissions and wastes) and unit costs, which are used in TEA and LCA (discussed in the following section).

### 2.1.3. Performing TEA and LCA

With the established *System* objects, cost analysis can be performed through the *TEA* class with additional user inputs (e.g., income tax, discount rate), and the costs will be updated each time the *System* object is simulated. Similarly, LCA is performed through the *LCA* class with the auto-generated inventory from *System*, but two more classes – *ImpactIndicator* and *ImpactItem* – are needed in using the *LCA* class. *ImpactIndicator* carries information on the impact indicators of interest (e.g., global warming potential). These indicators are then stored as attributes of *ImpactItem* objects, and these *ImpactItem* objects are used to represent inventory items such as construction materials, chemical inputs, and waste emissions. For each of the added indicators, *ImpactItem* objects also store the corresponding values of characterization factors (per functional unit of the impact item) provided by the users. *ImpactItem* can be linked to *WasteStream* and *SanUnit*, thereby enabling automatic updates of impact item quantities (e.g., $CO_2$ emitted during operation, cement required in construction) upon system simulation. In addition, *ImpactItem* objects can also be created in isolation with user-defined functions for automatic updates of item quantity upon simulation (e.g., system-wise electricity usage independent of process modeling but estimated empirically based on system operation time). Similar to system design and process simulation, results of TEA and LCA (e.g., total and breakdown of costs and environmental impacts) are accessible in the Python environment for further processing and can be output as data spreadsheets.

**2.1.4. Executing uncertainty and sensitivity analyses**

Finally, *Model* objects created in association with the *System* object of interest can be used to incorporate uncertainties in the system's design, simulation, TEA, and LCA. Two core attributes of the *Model* class – *Parameter* and *Metric* – are used to specify the input parameters with uncertainty (e.g., a component's nitrogen content, kinetic parameters of a process model, retention time of a reactor, chemical price, impact item characterization factor) and the output variables to evaluate (e.g., effluent quality, total cost, environmental impacts), respectively, in Monte Carlo simulations. To perform an uncertainty analysis, *Model* first generates a sample matrix from the probability distributions (e.g., uniform, triangular) as defined via the *Parameter* objects using the sampling method selected by the user (e.g., random,[37] Latin hypercube[38]). Then simulation, TEA, and LCA of the *System* object are carried out for each sample and the result metrics (defined by the user via *Metric* objects) are recorded.

Further, leveraging external Python libraries (e.g., SALib,[39] Matplotlib,[40] seaborn[41]), QSDsan also includes a *stats* module with a wide range of global sensitivity analysis methods (e.g., Spearman rank correlation, Morris one-at-a-time technique,[42] the Sobol method[43]) and visualization functions. All input parameters (i.e., sample matrix), output results (i.e., metrics), and generated figures can be accessed in Python or saved externally for additional processing.

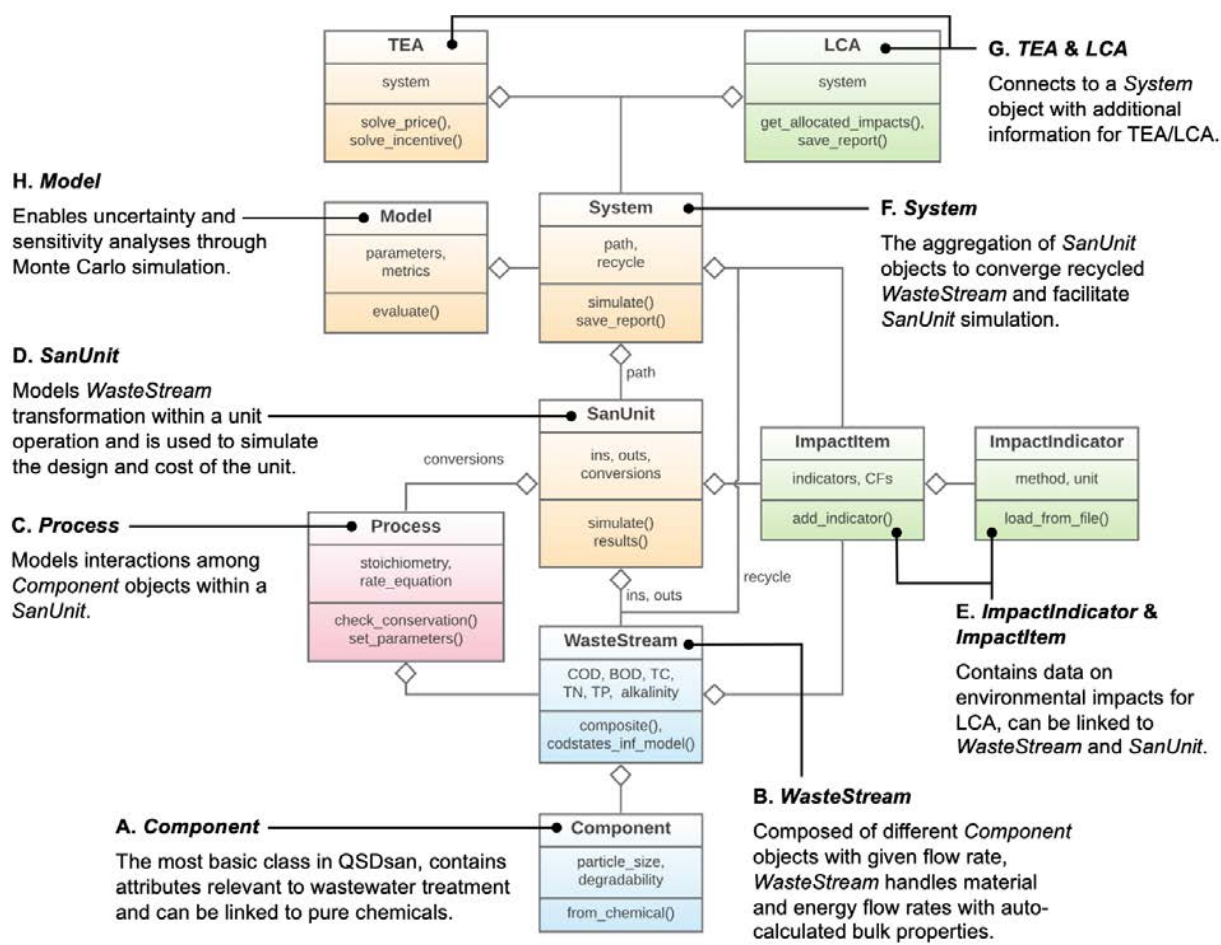

**Figure 1.** Simplified unified modeling language (UML) diagram showing the structure and core Python classes implemented in QSDsan. Each class is represented by a box containing the class name (bold, top part of the box) with select main data (middle part of the box) and methods (end with parentheses, bottom part of the box) as attributes. Letters **A**-**H** represent the class hierarchy from lower (i.e., more fundamental) to higher (i.e., more advanced) levels. The *Component* and *WasteStream* classes in blue are inherited from the *Chemical* and *Stream* classes in Thermosteam[33,34] (BioSTEAM's thermodynamic engine[31,32]) with the addition of wastewater-related attributes (e.g., ones noted in the box). The *Process* class in red enables dynamic simulation of *Component* objects' transformation during processes (e.g., degradation of substrates). The *SanUnit* class in yellow is inherited from the *Unit* class in BioSTEAM with added

capacity for dynamic simulation and handling of construction inventories, while the remaining yellow boxes (*System*, *TEA*, and *Model*) are imported from BioSTEAM. Green boxes including *ImpactItem*, *ImpactIndicator*, and *LCA* are implemented in QSDsan to enable LCA functionalities (explained in **Section 2.1.3**).

*2.2. Illustration of QSDsan Applications*

**2.2.1. Evaluation of a complete sanitation value chain**

To illustrate QSDsan's capacity in system design, simulation, TEA, and LCA, a complete sanitation value chain with three alternative systems was implemented using QSDsan. Details on the three systems can be found in Trimmer et al. (**Figure 2**, top panel).[25] Briefly, all three systems included user interface, onsite storage, conveyance, centralized treatment, and reuse units. In System A, a pit latrine was used as the user interface and onsite storage, followed by tanker truck transport to the existing treatment plant (sedimentation, anaerobic lagoon, facultative lagoon, and unplanted drying bed), and the recovered nutrients (N, P, and K) were sold as liquid fertilizers. For System B, the same pit latrine and tank trucker were used, but an anaerobic treatment plant (anaerobic baffled reactor, liquid treatment bed, and unplanted and planted drying beds) was modeled. Biogas can be recovered from the anaerobic baffled reactor and was assumed to be sold as a cooking fuel (as a replacement for liquid petroleum gas). Lastly, container-based toilet (urine-diverting dry toilet, UDDT), storage (urine storage tank and feces dehydration vault), and conveyance (handcart and tanker truck) facilities were designed for System C. For treatment and reuse, the same treatment and reuse processes as in System A was used, with the only exception being that the sedimentation unit was eliminated because solids were already separated from liquid in the UDDT. Solids recovered from all three systems were assumed to be used for land application.

The assumptions in design, TEA, LCA, and uncertainty parameters of the three systems followed those described in Trimmer et al.,[25] and the data spreadsheets are available to the public.[26] Six indicators were included in the analyses: total recoveries of COD, N, P, and K, as well as annual cost and emission. The recoveries are reported as the percentage of the excreted COD or nutrients. Annual cost is reported as $·cap$^{-1}$·yr$^{-1}$, and greenhouse gas emission is reported as global warming potential (GWP) in kg CO$_2$eq·cap$^{-1}$·yr$^{-1}$ using the TRACI method for life cycle impact assessment. A total of 137, 133, and 122 parameters were varied for Systems A, B, and C, respectively. Monte Carlo simulation of 5,000 runs using the Latin hypercube method to generate samples were performed for all systems. To enable pair-wise comparison of results in the uncertainty analysis, in each sample, a parameter was assigned the same value across any systems that share this parameter (e.g., the same pit latrine emptying time was used for both Systems A and B, the same user caloric intake was used for all systems).

To identify the key drivers of system sustainability, a global sensitivity analysis using the Morris one-at-a-time technique[42] was performed with 50 trajectories. Each trajectory represents one set of simulations that yield one evaluation of each parameter's "elementary effect" on the model outputs (e.g., the change in annual cost of the system caused by a change of pit latrine emptying period with other parameters fixed at certain values), and the total number of simulations for one system ($N_{Morris}$) is:

$$N_{Morris} = n_{trajectory} \times (k + 1) \quad \textbf{(Eqn. 1)}$$

where $n_{trajectory}$ is the number of trajectories and $k$ is number of parameters with uncertainty. Therefore, 6,900, 6,700, and 6,150 simulations were performed for Systems A, B, and C, respectively. Results of the Morris analysis were reported as µ* and σ values, which indicate the mean and the variance, respectively, of the evaluated "elementary effects" across trajectories. The µ* and σ values of each parameter were normalized by the largest µ* value of all input parameters for a particular indicator (e.g., cost) of one system so that the relative distribution of parameters in the µ*-σ plane across the six indicators in the three systems can be presented on the same scale.

Uncertainty and sensitivity analyses were performed and visualized using the *stats* module in QSDsan (except for minor annotation of the QSDsan-generated figures to improve readability). Source codes of the three systems, scripts used to conduct uncertainty and sensitivity analyses and plotting, results and figures (generated with QSDsan v1.1.3 and EXPOsan v1.1.4), and a brief instruction, can be found in the online repository EXPOsan (central location for systems developed using QSDsan).[44]

### 2.2.2. Process model benchmarking

Toward dynamic process modeling, the BSM1 system[27] (**Figure 4A**) was modeled with QSDsan following the MATLAB/Simulink implementation.[28,45] A *WasteStream* object representing a constant influent was created as a feed to the system, which is composed of a five-compartment activated sludge reactor modeled as five continuous stirred tank reactors (CSTR) in series followed by a secondary clarifier. Two *SanUnit* subclasses, representing CSTRs and flat-bottom circular clarifiers, respectively, have been included in QSDsan's portfolio of unit operations. Eight *Process* objects were created to describe ASM1's biokinetic processes[29] in all five CSTRs, the first two of which were anoxic. Two additional *Process* objects were created to represent the diffused aeration processes in the remaining three aerobic CSTRs ($K_La$ = 240 $d^{-1}$ for the first two aerobic CSTRs, O1 and O2, and $K_La$ = 84 $d^{-1}$ for the last, O3). A simple one-dimensional 10-layer settling model[30] was built into the clarifier to model particulate components, whereas soluble components were modeled as if in a non-active CSTR (i.e., one layer). Return activated sludge flowrate was controlled to be identical to the system influent flowrate. The return effluent flowrate from the last to the first CSTR was controlled by an effluent split ratio of 0.6, translating to an internal recirculation rate of 3 times the influent volumetric flow.

To verify the dynamic simulation algorithm in QSDsan, time-series data of state variables generated from simulation of the BSM1 system in QSDsan was compared against the IWA MATLAB/Simulink implementation[28] under identical conditions (constant influent, open-loop configuration; referred to as "baseline" hereinafter), and a simulation period of 50 day was selected where steady state could be achieved for both implementations. Complete scripts of BSM1 implementation in QSDsan, result data and figures (generated with QSDsan v1.1.3 and EXPOsan v1.1.4), and a brief instruction, can be found in the EXPOsan repository.[44] In addition, to test the rigor of QSDsan's algorithmic implementation of process models, the BSM1 system was simulated with 100 different initial conditions generated through Latin hypercube sampling from uniform distributions centered around the baseline initial concentrations (±50%). A complete Monte Carlo simulation (N=1,000) with 28 uncertain parameters (including 7 design and operational control decision variables) and 7 metrics was performed, followed by Monte Carlo filtering to identify the top two influential decision variables affecting effluent quality. An additional round of simulations varying the two most influential decision variables was performed to demonstrate QSDsan's capability to visualize system performance across the decision space. Specifics of simulation settings for the analyses above can be found in the supplementary material.

## 3. Results and Discussion

### 3.1. Groundwork toward an Open and Community-Led Platform.

The core structure of QSDsan has been completed and released on the Python Package Index (PyPI) repository.[46] All of QSDsan's source codes and documentation are available online.[46,47] Tutorials with step-by-step instructions from the installation of Python to the use of QSDsan have also been included in the documentation, and a complementary YouTube channel has been created with videos of QSD Group members demonstrating through the tutorials.[48]

Continuous development and maintenance of QSDsan will be supported by members of the QSD Group with contributions from the community, and we adhere to the Contributor Covenant Code of Conduct[49] for a just and equitable community. To encourage external contributions, QSDsan's documentation also includes a special section on contribution instructions and guidelines.[50] Briefly, the first contribution from a community developer will be on that developer's own "fork" (i.e., copy) of QSDsan, and the developer can submit a "pull request" to QSDsan's root repository (hosted by the QSD group) to merge the changes into QSDsan's stable version. The pull request will be accepted if the developer's fork (i) contains meaningful contributions with documentation, (ii) has no conflicts with the root repository, and (iii) has passed all test modules within QSDsan's core classes. After the first contribution, the developer will be invited to join the QSD group and given writing access to the root repository.

### *3.2. Simulating a Complete Sanitation Value Chain*

#### 3.2.1. Uncertainty analysis of sanitation alternatives

QSDsan can be used to simulate complete sanitation value chains and characterize their sustainability via TEA and LCA, thereby providing guidance when navigating tradeoffs among engineering performance metrics (e.g., nutrient recoveries) and sustainability indicators (e.g., cost, emissions) among alternative systems (**Figure 2**, middle and bottom panels). For nutrient recoveries, all three systems were able to recover the majority of the K in the excreta. System C was able to recover much more N than Systems A and B due to the separation of urine in UDDT. In contrast, System B achieved the highest P recovery as it avoided the loss of P to settled solids in System A's treatment processes and the precipitation of P (as struvite) in System C's decentralized urine storage unit. For COD recovery, Systems B and C were able to recover more COD due to the generation and capturing of biogas (System B) or less degradation of organics in the sludge (System C). As for cost and GWP, System B was the most affordable system because of the relatively inexpensive facilities and the revenue from selling biogas, while System C had the highest user cost due to the more expensive UDDT and the higher operating cost (more frequent waste collection). However, GWP of System C was the lowest among all as a result of the mitigation of fugitive emissions (less CH$_4$ and N$_2$O from organic degradation) and the offset of commercial fertilizers by recovered nutrients. Overall, these results are consistent with the previously published analysis (a summary spreadsheet including full comparison of the results is available through the online EXPOsan repository[44]), validating the core functionality of QSDsan.[25]

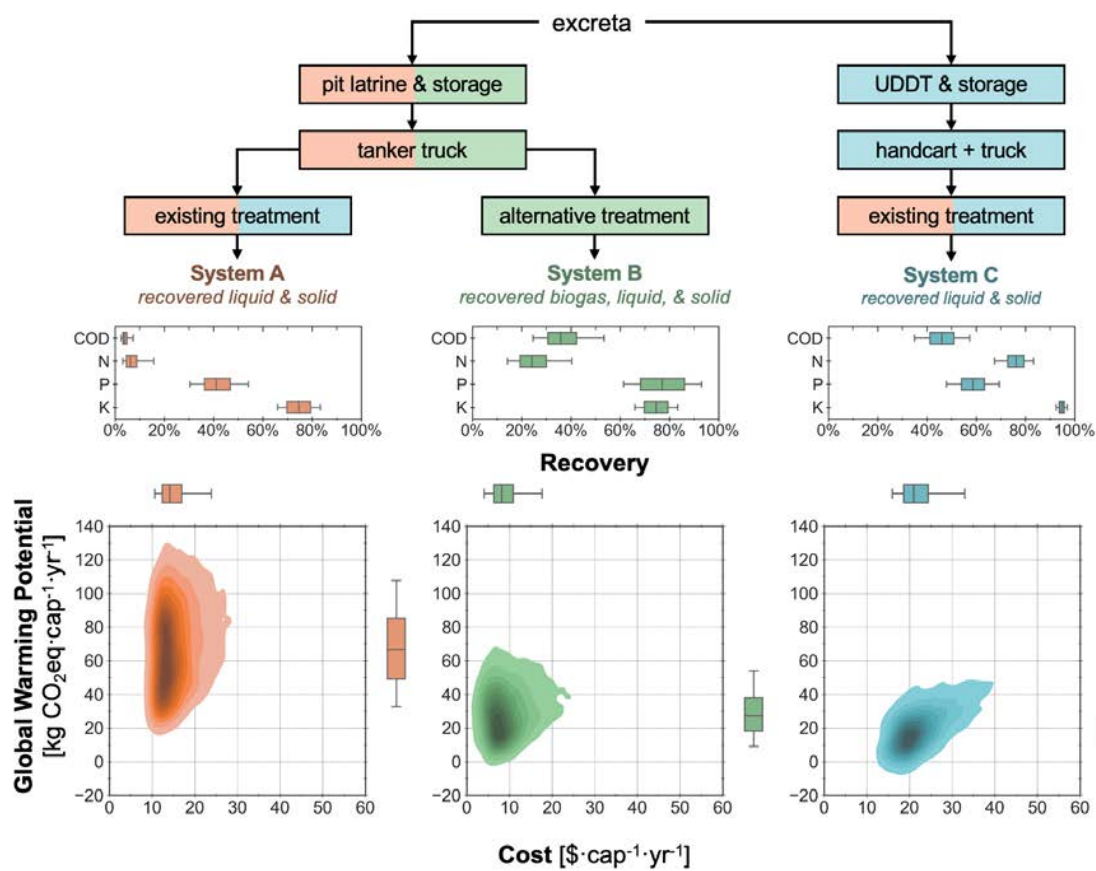

**Figure 2.** (**Top**) Simplified process diagrams, (**middle**) box plots of COD and nutrient (N, P, and K) recoveries expressed as percentage of the initial input in the excreta, and (**bottom**) kernel density plots of net annual costs and annual global warming potentials for the three sanitation systems as described in Trimmer et al.[25] For the box plots in the middle and bottom panels, middle line, edges, and whiskers and are 50$^{th}$, 25$^{th}$/75$^{th}$, and 5$^{th}$/95$^{th}$ percentiles, respectively. Systems A, B, and C are color-coded in orange, green, and blue, respectively, in all panels.

**3.2.2. Global sensitivity analysis of sanitation alternatives**

The built-in global sensitivity analyses in QSDsan can be leveraged to identify drivers of system sustainability and understand the interactions between uncertain parameters. In this illustration, the Morris technique was chosen because of (i) its effectiveness toward screening the most impactful parameters from models with many uncertain parameters (137, 133, and 122 parameters for Systems A, B, and C, respectively) at relatively small sample sizes; (ii) its ability to provide insights on the interactions between parameters; and (iii) its applicability to nonlinear and non-monotonic system models. Results from the Morris analysis are commonly reported as two sensitivity indices, $\mu^*$ and $\sigma$. For a particular indicator, parameters with larger $\mu^*$ values are considered to have a more significant effect on the indicator than other parameters. A larger $\sigma$ indicates the parameter having a nonlinear effect on the results and/or stronger interactions with other parameters (i.e., the effect of this parameter on the result has a stronger dependency on the values of other parameters). Moreover, a parameter with a $\sigma$-to-$\mu^*$ ratio ($\sigma/\mu^*$) greater than 1 is usually considered as having a non-monotonic effect on the results, whereas a parameter with $\sigma/\mu^*<0.1$ is considered as having an almost linear effect on the results.[51,52]

Results from the Morris analysis revealed different trends of the effect of key parameters on systems and indicators (**Figure 3** and Figures S1-S2 in the supplementary material). For COD and N, recoveries for Systems A and B were controlled by parameters associated with the pit latrine, and these parameters were found to have stronger interactions (i.e., larger $\sigma/\mu^*$ values) than those of System C. In Systems A and B, the parameter pit latrine emptying period (i.e., the time between two emptying events) was the most significant driver for COD and N recoveries, but its impacts were realized in combination with other parameters (e.g., maximum degradation, the time to reach full degradation, the reduction at full degradation). For System C, however, because of the much less degradation in UDDT, the recoveries of COD and N were controlled by parameters associated with different units within the system, and their effects were less prominent. For P, key parameters for Systems A and B were related to leaching or the treatment processes (e.g., P retention in the sedimentation tank, P removal in the lagoon), whereas for System C, dietary parameters (e.g., P intake, P and Ca contents in urine) were critical due to the potential precipitation reactions in the urine storage tank that could lead to the loss of P. In the case of K, due to its better retention through the systems (as observed in the uncertainty analysis), its recovery was mostly sensitive to the amount lost to leaching (for pit latrines in Systems A and B) or handling during conveyance and application of the recovered nutrients (for System C).

Regarding the annual cost, household size had the largest normalized $\mu^*$ and $\sigma$ values compared to other parameters across all systems, indicating that it was the main driver of the overall system cost, and its impacts on the cost were realized in combination with other parameters. This is because the household size directly determined the number of toilets needed for the system, which affect the costs and emissions of not only the user interface units (i.e., pit latrine or UDDT) but also downstream units (e.g., storage and conveyance units). As the number of toilets was also dependent on the total population and household toilet use density (number of households served by a toilet), a larger interaction effect was observed (i.e., larger $\sigma$ value).

For GWP, household size remained the most impactful parameter for System C. However, for Systems A and B, the percentage of caloric intake diverted to the excreta had the largest effect as it contributed to direct $CH_4$ emission (from the degraded organic matter in the excreta), which was the largest contributor to GWP as observed in the uncertainty analysis.

Finally, Systems A and B (same toilet, storage, and conveyance units, different treatment processes) had much more similar patterns of $\mu^*$ and $\sigma$ (e.g., $\sigma/\mu^*$ values, key parameters) than Systems A and C (same treatment process, different toilet, storage, and conveyance units), revealing that the selection of toilet (and therefore storage and conveyance units) was more impactful to the sustainability of the sanitation value chain than the choice of treatment process. Overall, this example illustrates QSDsan's capacity in system design, simulation, TEA, LCA, as well as the utility of its statistical module in carrying out integrated uncertainty and sensitivity analyses and providing visualization tools to guide the RD&D of technologies.

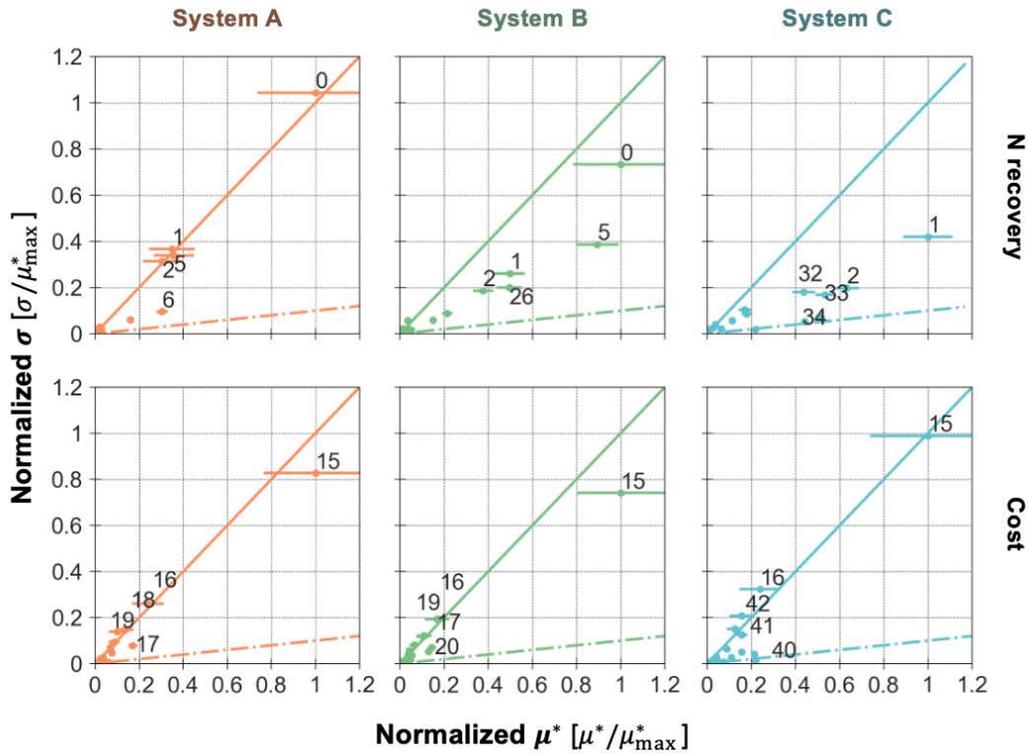
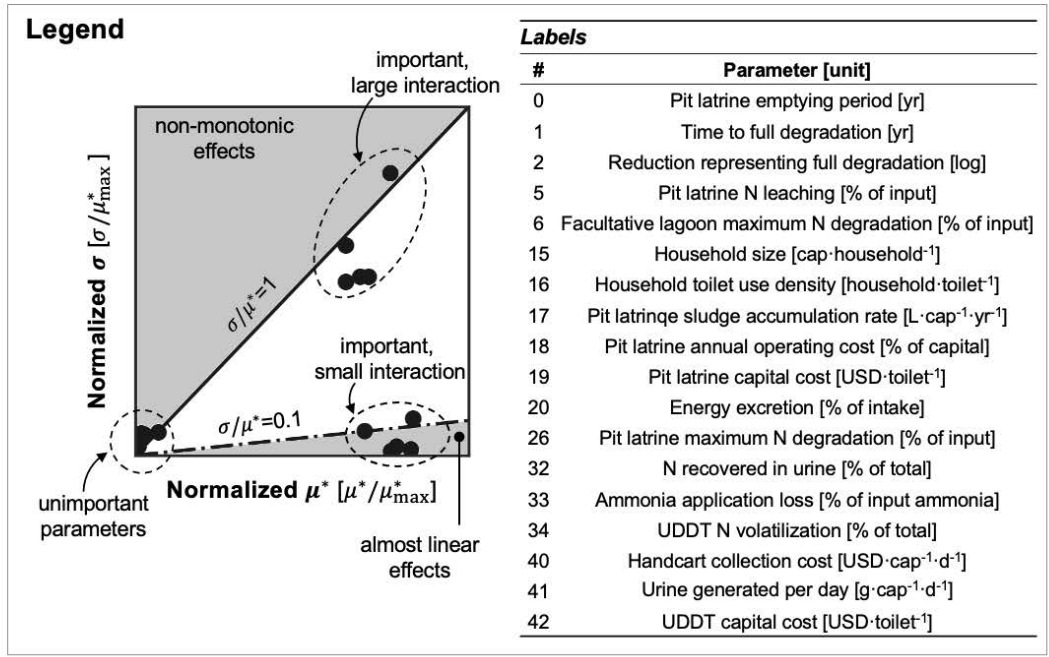

**Figure 3.** Morris sensitivity indices (μ* and σ) of the key parameters for N recovery and cost of the three systems. Each plot represents the indices calculated for one indicator of a system. Error bars represent the 95% confidence interval of μ*. For illustration purpose (i.e., use the same x

and y axes scales across different indicators and systems), values of µ* (x-axis) and σ (y-axis) were normalized by the maximum µ* of all parameters in the analyzed system for the analyzed indicator. Key parameters were defined as parameters with a normalized µ* value greater than or equal to 0.1 ($\mu^*/\mu^*_{max} \geq 0.1$). If there were more than five parameters meeting this criterion, only the top five parameters with the highest normalized µ* values were considered key parameters. All parameters were included in the plot, but only the key parameters were labeled (parameters with small normalized µ* and σ values clustered at the origin point are thus indistinguishable). In all plots, the solid and dashed lines have slopes of 1 and 0.1 (i.e., σ/µ* = 1 and 0.1), respectively, where the parameters on the left side of the solid line were considered to have non-monotonic effects on the indicator values, and parameters on the right side of the dashed line were considered to have linear effects. A compiled figure including all indicators is included as Figure S1 (scatter plots as this figure) and S2 (bubble plot) in the supplementary material; all generated raw data and figures are available online.[44]

### 3.3. Benchmarking a Pseudo-Mechanistic Process Model

#### 3.3.1. Dynamic simulation of BSM1

Simulation results of BSM1 show that QSDsan is capable of implementing complex process models correctly and running dynamic simulations at speeds comparable to existing process simulation options (**Figure 4B**). On a personal computer with an Intel Core i7-6700K CPU @ 4.00 GHz and 16.0 GB RAM, it took about 4 to 6 seconds on average for QSDsan to initialize system state, compile ODEs, and perform a 50-day simulation of BSM1 with any implicit ODE solver readily available in the SciPy package.[22] Regardless of the initial conditions assigned to each simulation, the system consistently converged to the same steady state in QSDsan, which matched results from MATLAB/Simulink with a maximum relative error < 1% for state variables across all unit operations in the system.

The transparent implementation of BSM1 provides an illustrative example of how QSDsan's basic structure can be leveraged and built upon for a wide range of process modeling applications. Users can utilize existing scripts of activated sludge models as templates for implementing new process models developed for novel technologies (e.g., microalgal and cyanobacterial process models for photobioreactors[53]). New unit operations can be added as subclasses of *SanUnit* to the existing portfolio for dynamic simulations with essential attributes (e.g., ODE algorithms) describing the mass balance. QSDsan's current structure also supports the implementation of more complex simulation settings, such as dynamic influent streams and active operational control with proportional–integral–derivative (PID) controllers, the corresponding subclasses of which are under active development.

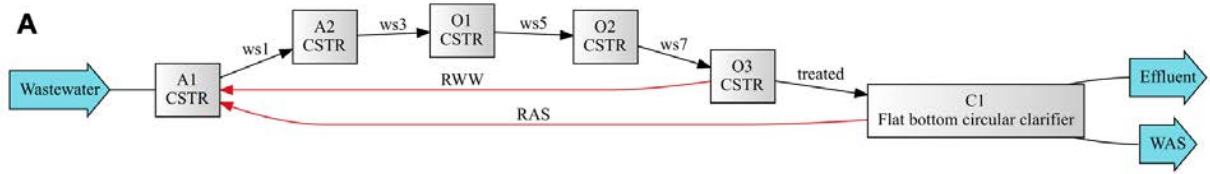

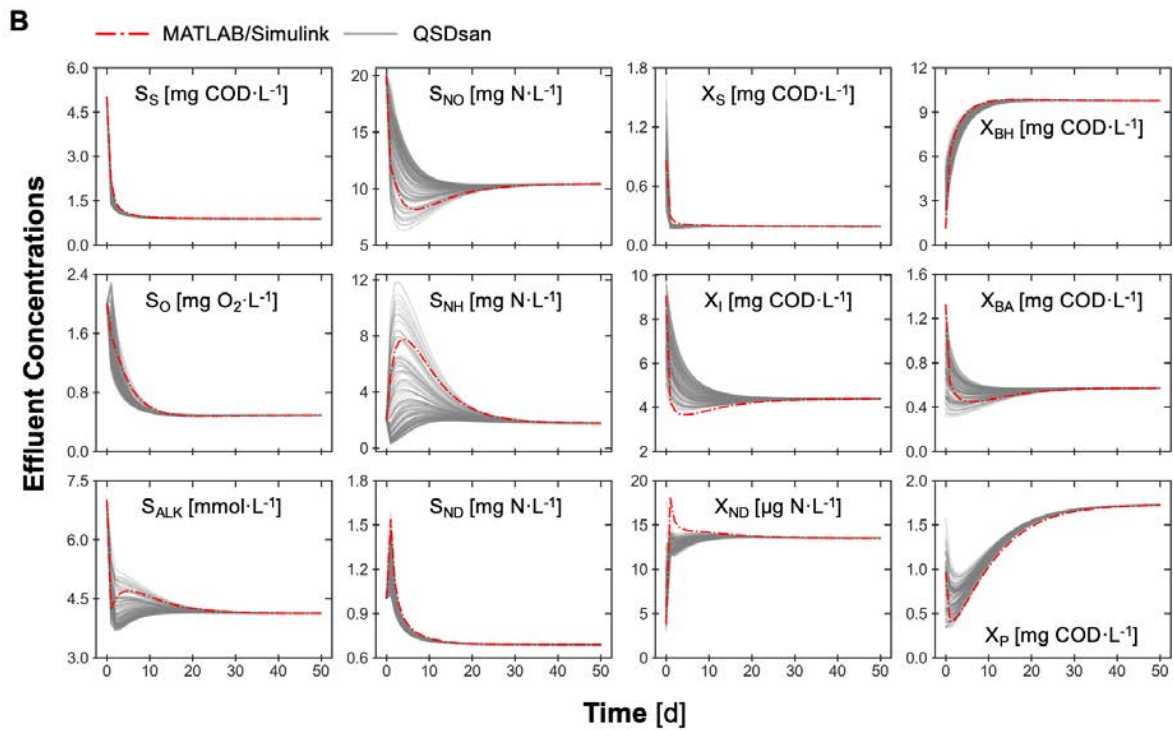

**Figure 4.** (**A**) Configuration of the BSM1 system generated by QSDsan. A1 and A2 represent the anoxic reactors; O1, O2, and O3 represent the aerated reactors. RWW – return wastewater; RAS – return activated sludge; WAS – waste activated sludge. (**B**) Dynamics of BSM1 system's effluent state variables simulated by QSDsan (with 100 different initial conditions) and MATLAB/Simulink (with the baseline initial condition). Soluble inert organic matters, $S_I$, is omitted in this figure because it is not involved in any conversion processes, as defined in ASM1.

### 3.3.2. Uncertainty and sensitivity analyses of BSM1

Uncertainty in technological parameters (e.g., heterotrophic yield in ASM1) and contextual parameters (e.g., influent ammonium concentration, saturation dissolved oxygen [DO]), as well as changes in decision variables (e.g., aeration flowrate, designed reactor volume) and modeling assumptions (e.g., COD-to-mass ratio of particulate organic substrates) can be simultaneously incorporated into the system and propagated through Monte Carlo simulation to characterize the uncertainty in system model outputs. For example, visualization of the effluent dynamics over time and the converged steady state (**Figure 5**) allows users to quickly identify that state variables closely related to the nitrification/denitrification processes (i.e., nitrite/nitrate $S_{NO}$, and ammonia $S_{NH}$) were subject to the greatest uncertainty. Similarly, by inspecting the distributions of key metrics (common effluent quality indicators, daily sludge production, sludge retention time [SRT]) against control or design objectives (e.g., regulations on effluent quality, or sludge disposal costs), one can locate the performance gaps and make targeted decisions for improvement. For example, assuming the discharge limits are TN ≤ 18 mg-N·L$^{-1}$ and TKN ≤ 4.0 mg-N·L$^{-1}$, the BSM1 system was estimated to have 11.3% and 43.3% chances of violation, respectively, while the discharge limits of COD, BOD$_5$, and TSS (assumed to be 100 mg·L$^{-1}$, 10 mg·L$^{-1}$, and 30 mg·L$^{-1}$, respectively) were estimated not binding for the BSM1 system (Figure S3 in the supplementary material).[27]

To identify the driving factors for compliance with the discharge limits (Table S4), Monte Carlo filtering of the 28 parameters was performed with the simulation data from the uncertainty analysis above. By dividing the samples into two groups (i.e., above and below the discharge limit of TN) and performing Kolmogorov–Smirnov tests to compare the distributions of each parameter between groups, we found the aeration rate to the first two aerobic CSTRs and the aerobic zone hydraulic retention time (HRT) have the most significant differences in distribution between the two groups and thus identified them as the most impactful decision variables for effluent TN (Table S4). For effluent TKN, in addition to the decision variables above, waste activated sludge (WAS) flowrate and return sludge flowrate, were also found to have significant impacts on the compliance with the discharge limit. These decision variables directly affect the DO levels and the SRT of the activated sludge system, which have been commonly recognized among the most important operational control and design parameters for biological nitrogen removal.[54,55] This shows QSDsan is equipped with robust algorithms for process simulation and uncertainty and sensitivity analyses, demonstrating its capacity for systematic identification of key decision variables and impactful parameters.

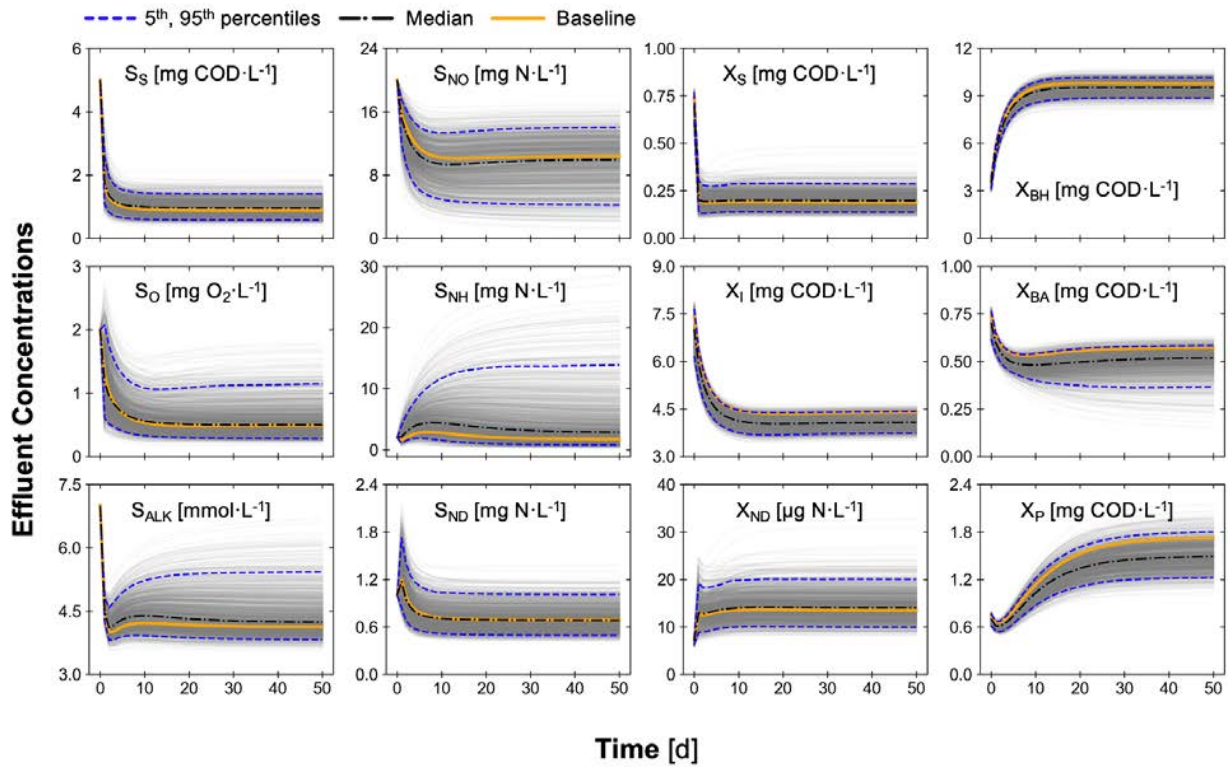

**Figure 5.** Dynamics of the effluent state variables generated in Monte Carlo simulations of the BSM1 system for uncertainty analysis. Soluble inert organic matter, $S_I$, is omitted in this figure because it is not involved in any conversion processes, as defined in ASM1. Each solid grey line represents one sample. Median and 5th/95th percentiles at each time point were calculated based on the entire 1,000 samples.

### 3.3.3. Mapping the decision space of activated sludge system operation

With the most impactful decision variables identified, systems can be simulated in QSDsan across the decision space of these variables to elucidate their implications on system performance (**Figure 6**). Based on the results from Monte Carlo filtering, WAS flowrate and aeration flowrate to the first two aerobic CSTRs were chosen to illustrate the quantitative investigation of an optimal control strategy for the BSM1 system with QSDsan. Aerobic zone HRT, albeit important for effluent TN and TKN, was excluded from the analysis as it would be determined upon the design of the reactors. Results showed that effluent COD and $BOD_5$ were insensitive to the change of sludge wastage flowrate, whereas varying aeration rate had little impact on effluent TSS. Moderate reduction of both WAS and aeration flowrates from the baseline levels had the benefits of further lowering effluent TN and daily sludge production with only a marginal increase of effluent TSS and COD. This observation implies that the system can be operated at lower aeration energy demand and sludge disposal cost while meeting the discharge regulatory requirements. With the *TEA* and *LCA* classes in QSDsan, economic and environmental implications of such operational control changes can be further quantified and leveraged to understand potential trade-offs (e.g., reducing cost vs. greater risk of violating limits).

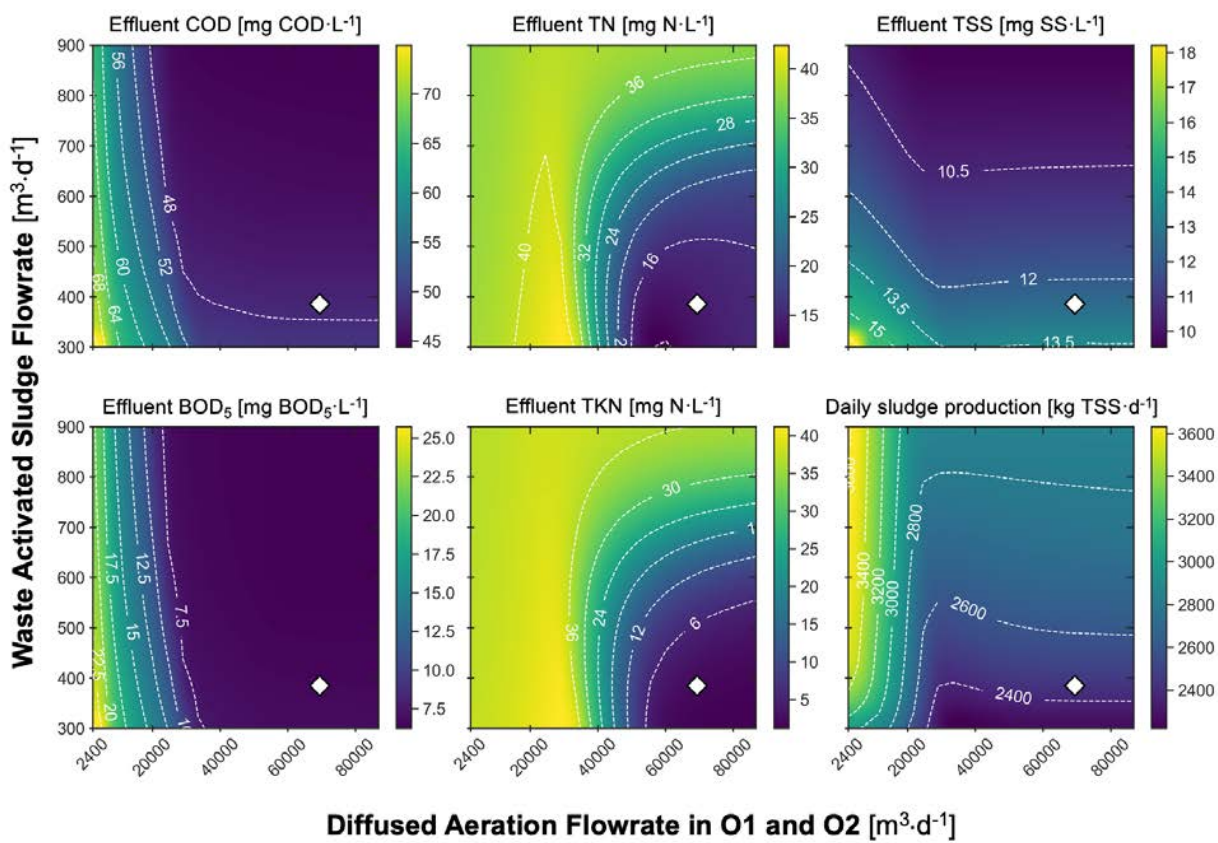

**Figure 6.** Effluent quality metrics mapped across the decision space of aeration and sludge wastage. The diffused aeration air flowrate at field condition in the first two aerated CSTRs (i.e., O1 and O2) was varied uniformly between 2,400 $m^3 \cdot d^{-1}$ and 86,765 $m^3 \cdot d^{-1}$, corresponding to a $K_La$

of 80-300 d$^{-1}$. The WAS flowrate was varied uniformly within 300-900 m$^3$·d$^{-1}$, resulting in an SRT of 4.15-17.2 days. The white diamonds represent the baseline setting.

*3.4. Future Work Enabled by QSDsan*

With diverse and growing capacities in system design, simulation, and sustainability analyses including TEA and LCA, QSDsan can be used to prioritize research directions, facilitate technology deployment, and navigate decision-making across wide ranges of decision, technological, and contextual parameters. In particular, the OOP paradigm allows users to utilize existing units and systems, tailor the units and systems to their needs (e.g., adjust cost calculation), and/or develop new units and systems that suit their needs. Uncertainties in every element of the design and operation (e.g., material costs, technology performance) can be included in system simulation and sustainability analyses. This flexibility in system design and ability to perform rigorous uncertainty and sensitivity analyses (especially global sensitivity analysis) are critical to emerging technologies that are characterized by large uncertainties in their design and performance.

Further, the agility of QSDsan allows it to be easily connected to external packages for enhanced features. For example, through BW2QSD[56] and Brightway2[57], users can directly retrieve life cycle inventory data from databases such as ecoinvent (requires ecoinvent license, example usage in EXPOsan[44]); through DMsan (decision-making for sanitation and resource recovery systems),[58] users can leverage QSDsan-generated simulation and sustainability analysis results in multi-criteria decision analysis. With Python being one of the most widely used programming languages (ranked #1 by IEEE Spectrum in 2021[59]), QSDsan will benefit from the rapidly growing number of Python modules and libraries for future improvement (e.g., incorporation of machine learning in mechanistic modeling,[60] implementation of digital twin in water/wastewater utilities[61]).

Moreover, by laying the groundwork (e.g., open-source, detailed documentation, executable tutorials with video demonstrations) for a collaborative platform, we aspire to build QSDsan as an open and community-led platform. This platform is poised to be adopted by researchers, practitioners (e.g., Container-Based Sanitation Alliance[62]), and the public across the world to increase access to and sustainability of sanitation, which is particularly relevant to resource-limited communities where the largest need is expected for the coming decades.[63] Additionally, QSDsan can be used in courses focusing on sustainable design to offer students hands-on opportunities to design their own systems and perform TEA and LCA without the cost barrier of a software license. Altogether, QSDsan provides the field of sanitation and resource recovery with a valuable and timely tool for guiding technology RD&D, informing decision making, and fostering the stewardship of sustainability among the next generation, ultimately contributing to the society's advancement toward a more sustainable future.

## 4. Conclusions

- QSDsan is an open-source platform that integrates system design, simulation, and sustainability characterization (TEA and LCA) under uncertainty for sanitation and resource recovery systems.
- Core capacities of QSDsan including tailorable design, flexible process modeling, rapid simulation, as well as advanced statistical analyses with integrated visualization.
- We illustrate the capacity of QSDsan through two examples: (i) equilibrium simulation, TEA, and LCA of three sanitation systems and (ii) dynamic simulation of the BSM1 system, both with robust uncertainty and global sensitivity analyses.
- QSDsan can be leveraged to prioritize research and inform decision-making, thereby supporting and expediting the RD&D of sanitation and resource recovery technologies.

- We strive to build QSDsan as a community-led platform with online documentation, tutorials (including video demonstrations), and transparent management with clear contribution guidelines.

**Acknowledgements**

This publication is based on research funded in part by the Bill & Melinda Gates Foundation. The findings and conclusions contained within are those of the authors and do not necessarily reflect positions or policies of the Bill & Melinda Gates Foundation. We would like to thank Dr. Ulf Jeppsson (Lund University) for providing advice on the Benchmark Simulation Model and Yoel Cortés-Peña (University of Illinois Urbana-Champaign) for the suggestions and discussions on package development.

Supplementary Material for

# QSDsan: An Integrated Platform for Quantitative Sustainable Design of Sanitation and Resource Recovery Systems

Yalin Li[a,b†]*, Xinyi Zhang[c†], Victoria L. Morgan[a], Hannah A.C. Lohman[c], Lewis S. Rowles[a,1], Smiti Mittal[d], Anna Kogler[e], Roland D. Cusick[c], William A. Tarpeh[e,f], Jeremy S. Guest[a,b,c]

[a] Institute for Sustainability, Energy, and Environment, University of Illinois Urbana-Champaign, 1101 W. Peabody Drive, Urbana, IL 61801, USA.

[b] DOE Center for Advanced Bioenergy and Bioproducts Innovation, University of Illinois Urbana-Champaign, 1206 W. Gregory Drive, Urbana, IL 61801, USA.

[c] Department of Civil and Environmental Engineering, University of Illinois Urbana-Champaign, 3221 Newmark Civil Engineering Laboratory, 205 N. Mathews Avenue, Urbana, IL 61801, USA.

[d] Department of Bioengineering, Stanford University, 129 Shriram Center, 443 Via Ortega, Stanford, California 94305, USA.

[e] Department of Civil and Environmental Engineering, Stanford University, 311 Y2E2, 473 Via Ortega, Stanford, California 94305, USA.

[f] Department of Chemical Engineering, Stanford University, 129 Shriram Center, 443 Via Ortega, Stanford, California 94305, USA.

[1] Present address: Civil Engineering and Construction, Georgia Southern University, 201 COBA Drive, BLDG 232 Statesboro, GA 30458, USA.

[†]**Co-first authors:** Y. Li and X. Zhang contributed equally to this work.

**\*Corresponding author:** yalinli2@illinois.edu, +1 (217) 418-4672

9 pages, 4 tables, 3 figures

## S1. Additional Morris Analysis Results of the Bwaise System

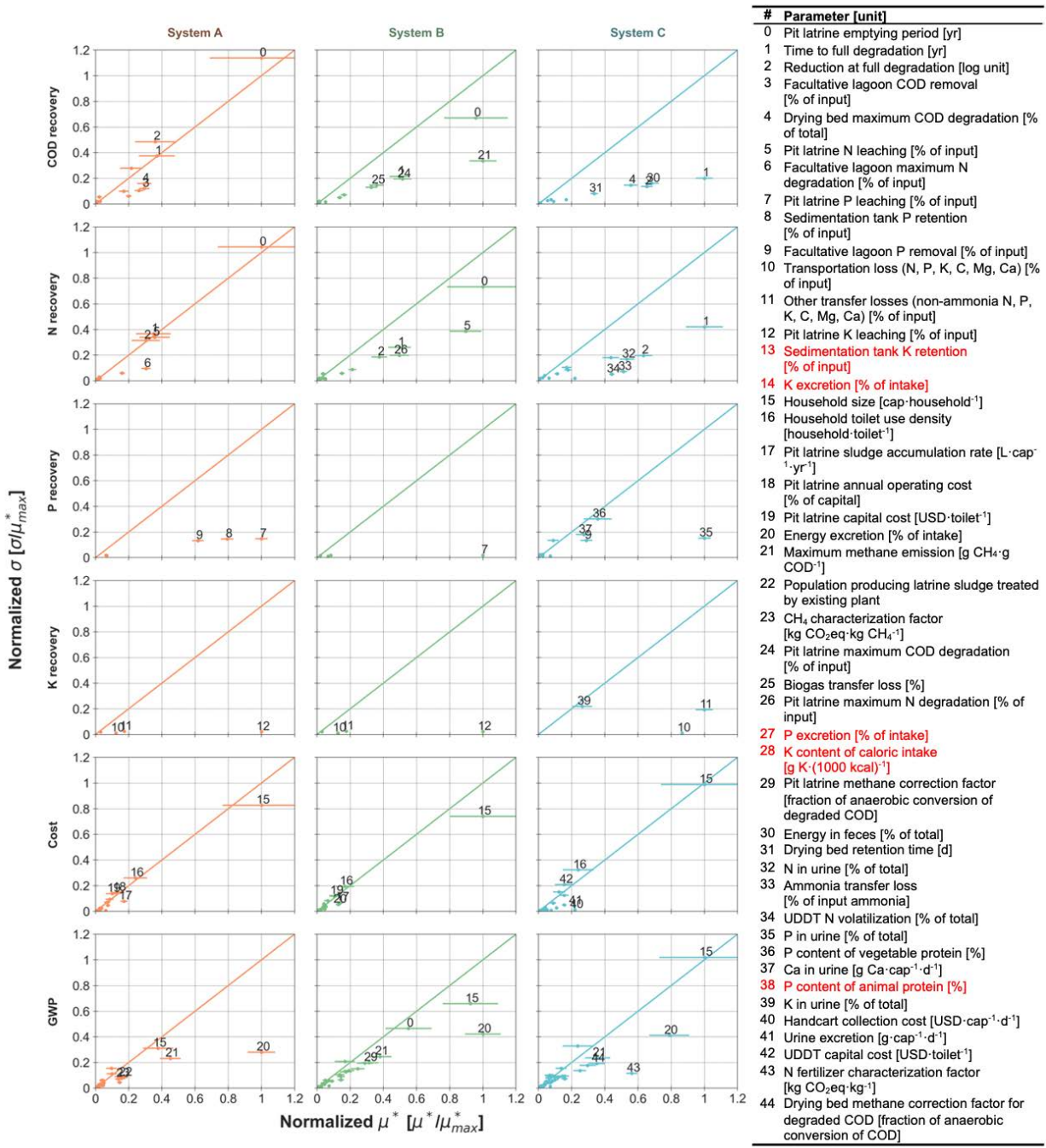

**Figure S1.** Complete figure of the Morris analysis results (refer to **Figure 3** in the main text for the full legend). Parameters in red fonts were one of the five parameters with the largest normalized µ* values for a given metric of the corresponding system, but the normalized µ* values were smaller than 0.1 and not considered as "key parameters". Raw data (including full list of the parameters and their µ* and σ values) can be found in the *bwaise* module (the cached_results_figures folder) of the EXPOsan repository online (https://github.com/QSD-Group/EXPOsan/tree/main/exposan/bwaise).

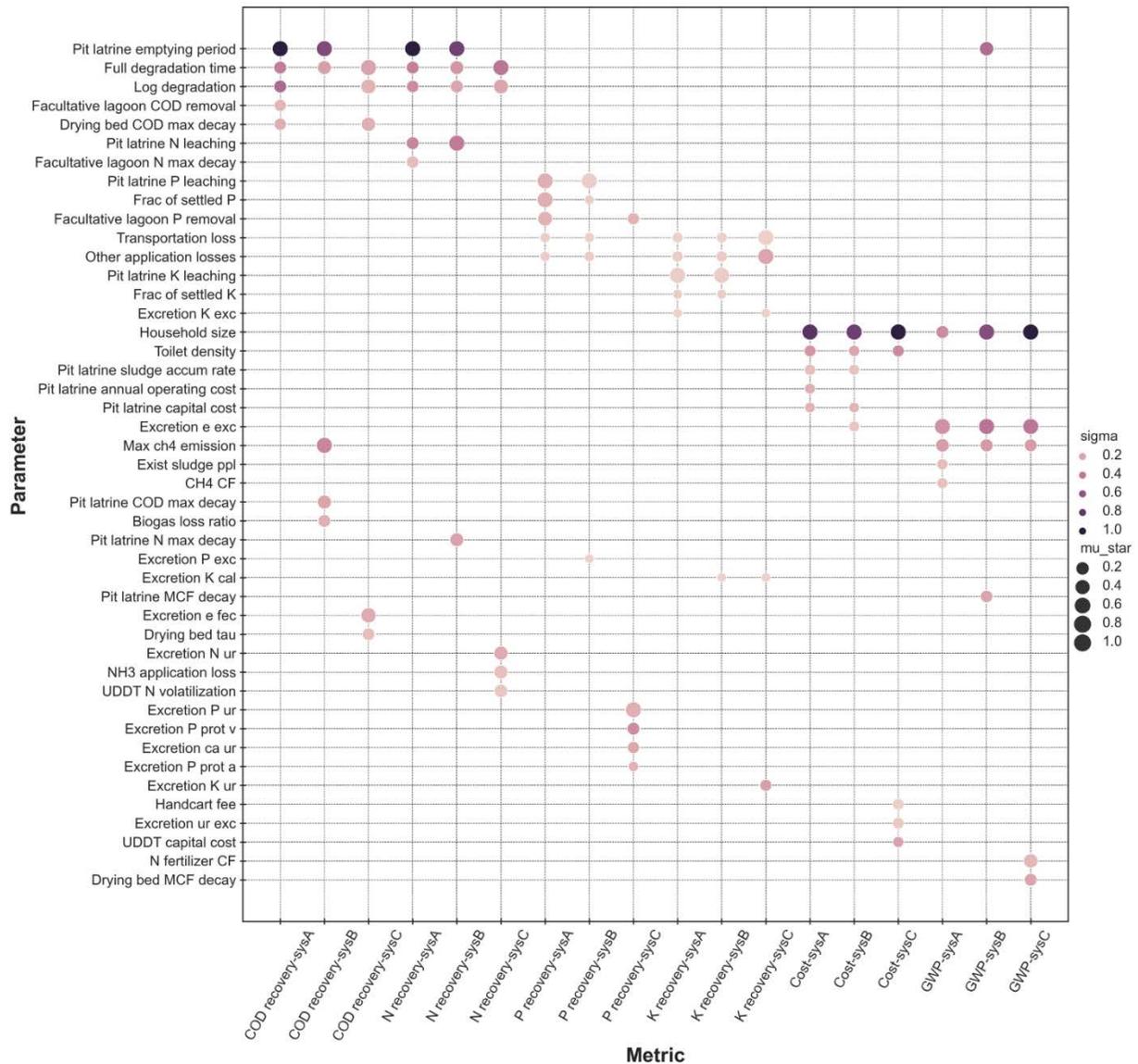

**Figure S2.** Alternative figure presenting the Morris analysis results (refer to Figure S1 for the name and unit of the parameters, parameters were listed in the same order as in the legend table in Figure S1). Raw data (including full list of the parameters and their μ* and σ values) can be found in the *bwaise* module (the cached_results_figures folder) of the EXPOsan repository online (https://github.com/QSD-Group/EXPOsan/tree/main/exposan/bwaise).

## S2. BSM1 Simulation Settings

### S2a. BSM1 system settings

**Table S1**. BSM1 system settings in baseline dynamic simulation and in uncertainty analysis. All uncertainty ranges were obtained from Sin et al.[1]

| Variable | | Description | Unit | Baseline | Uncertainty Analysis | | |
|---|---|---|---|---|---|---|---|
| | | | | | Minimum | Maximum | Distribution |
| Influent | $Q_{in}$ | Volumetric flowrate | m³·d⁻¹ | 18,446 | - | - | - |
| | $T_{water}$ | Water temperature | K | 293.15 | - | - | - |
| | $DO_{sat}$ | Saturation DO at field condition | mg-O$_2$·L⁻¹ | 8.0 | 7.2 | 8.8 | Uniform |
| | $S_S$ | Soluble organic substrate | mg-COD·L⁻¹ | 69.5 | - | - | - |
| | $X_{BH}$ | Active heterotrophic biomass | mg-COD·L⁻¹ | 28.17 | - | - | - |
| | $X_S$ | Particulate organic substrate | mg-COD·L⁻¹ | 202.32 | - | - | - |
| | $X_I$ | Particulate inert organic matter | mg-COD·L⁻¹ | 51.2 | - | - | - |
| | $S_{NH}$ | Ammonium nitrogen | mg-N·L⁻¹ | 31.56 | - | - | - |
| | $S_I$ | Soluble inert organic matter | mg-COD·L⁻¹ | 30 | - | - | - |
| | $S_{ND}$ | Soluble biodegradable organic nitrogen | mg-N·L⁻¹ | 6.95 | - | - | - |
| | $X_{ND}$ | Particulate biodegradable organic N | mg-N·L⁻¹ | 10.59 | - | - | - |
| | $S_{ALK}$ | Alkalinity, assumed to be bicarbonate | mmol·L⁻¹ | 7 | - | - | - |
| Environment | $T_{air}$ | Air temperature | K | 293.15 | - | - | - |
| | $P$ | Atmospheric pressure | Pa | 101,325 | - | - | - |
| Reactors | $V_a$ | Anoxic CSTR volume | m³ | 1,000 | 900 | 1,000 | Uniform |
| | $V_o$ | Aerobic CSTR volume | m³ | 1,333 | 1,200 | 1,333 | Uniform |
| | $K_La_1$ | Oxygen mass transfer coefficient at field condition for O1 and O2 reactors | d⁻¹ | 240 | 180 | 360 | Uniform |
| | $K_La_2$ | Oxygen mass transfer coefficient at field condition for O3 reactor | d⁻¹ | 84 | 75.6 | 92.4 | Uniform |
| | $H$ | Clarifier height | m | 4 | - | - | - |
| | $A$ | Clarifier surface area | m² | 1,500 | - | - | - |
| | $Q_{RAS}$ | RAS flowrate | m³·d⁻¹ | $1 \times Q_{in}$ | $0.75 \times Q_{in}$ | $1 \times Q_{in}$ | Uniform |
| | $Q_{WAS}$ | WAS flowrate | m³·d⁻¹ | 385 | 346.5 | 423.5 | Uniform |
| | $Q_{intr}$ | Internal recirculation flowrate | m³·d⁻¹ | $3 \times Q_{in}$ | $2.25 \times Q_{in}$ | $3.75 \times Q_{in}$ | Uniform |

## S2b. ASM1 parameters

**Table S2**. ASM1 parameters in baseline dynamic simulation and in uncertainty analysis. All uncertainty ranges were obtained from Sin et al.[1]

| Variable | | Description | Unit | Baseline | Uncertainty Analysis | | |
|---|---|---|---|---|---|---|---|
| | | | | | Minimum | Maximum | Distribution |
| Stoichiometry | $Y_H$ | Heterotrophic biomass yield on soluble substrate | g-COD·(g-COD)$^{-1}$ | 0.67 | 0.64 | 0.70 | Triangular |
| | $Y_A$ | Autotrophic biomass yield on ammonium N | g-COD·(g-N)$^{-1}$ | 0.24 | 0.23 | 0.25 | Triangular |
| | $f_{Pobs}$ | Observed fraction of inert particulate generated during biomass decay | unitless | 0.21 | 0.16 | 0.26 | Triangular |
| | $i_{XB}$ | Active biomass N content | g-N·(g-COD)$^{-1}$ | 0.08 | 0.04 | 0.12 | Triangular |
| | $i_{XP}$ | Cell product and inert particulate N content | g-N·(g-COD)$^{-1}$ | 0.06 | 0.057 | 0.063 | Triangular |
| | $f_{SS,COD}$ | mass-to-COD ratio of particulates | g·(g-COD)$^{-1}$ | 0.75 | 0.7 | 0.95 | Triangular |
| Kinetics | $\mu_H$ | Heterotrophic maximum specific growth rate | d$^{-1}$ | 4 | 3 | 5 | Triangular |
| | $K_S$ | Readily biodegradable substrate half saturation coefficient | g-COD·m$^{-3}$ | 10 | 5 | 15 | Triangular |
| | $K_{OH}$ | Oxygen half saturation coefficient for heterotrophic growth | g-O$_2$·m$^{-3}$ | 0.2 | 0.1 | 0.3 | Triangular |
| | $K_{NO}$ | Nitrate half saturation coefficient | g-N·m$^{-3}$ | 0.5 | 0.25 | 0.75 | Triangular |
| | $b_H$ | Heterotrophic biomass decay rate constant | d$^{-1}$ | 0.3 | 0.285 | 0.315 | Triangular |
| | $\mu_A$ | Autotrophic maximum specific growth rate | d$^{-1}$ | 0.5 | 0.475 | 0.525 | Triangular |
| | $K_{NH}$ | Ammonium (nutrient) half saturation coefficient | g-N·m$^{-3}$ | 1 | 0.5 | 1.5 | Triangular |
| | $K_{OA}$ | Oxygen half saturation coefficient for autotrophic growth | g-O$_2$·m$^{-3}$ | 0.4 | 0.3 | 0.5 | Triangular |
| | $b_A$ | Autotrophic biomass decay rate constant | d$^{-1}$ | 0.05 | 0.04 | 0.06 | Triangular |
| | $\eta_g$ | Reduction factor for anoxic growth of heterotrophs | unitless | 0.8 | 0.6 | 1.0 | Triangular |
| | $k_a$ | Ammonification rate constant | m$^3$·(g-COD)$^{-1}$·d$^{-1}$ | 0.05 | 0.03 | 0.08 | Triangular |
| | $k_h$ | Hydrolysis rate constant | d$^{-1}$ | 3 | 2.25 | 3.75 | Triangular |
| | $K_X$ | Slowly biodegradable substrate half saturation coefficient for hydrolysis | g-COD·(g-COD)$^{-1}$ | 0.1 | 0.075 | 0.125 | Triangular |
| | $\eta_h$ | Reduction factor for anoxic hydrolysis | unitless | 0.8 | 0.6 | 1.0 | Triangular |

## S2c. Initial conditions

The baseline initial conditions were used in both benchmarking dynamic simulation and the Monte Carlo simulations for uncertainty analysis. In steady-state convergence test, the five CSTRs shared identical initial conditions in each simulation, and the clarifier's initial soluble concentrations were set to equal its influent's concentrations at t=0. The clarifier's initial TSS in each layer, if not specified, were assumed proportional to influent TSS by fixed factors, which can be found in the *Clarifier* class of QSDsan. Per the assumption of the 1D 10-layer settling model[2,3], the compositions of particulates in clarifier influent are propagated immediately to its effluents.

**Table S3**. Initial conditions used in dynamic simulations and varied to test convergence of steady states.

| Variable | Description | Unit | Baseline CSTRs | Baseline Clarifier | Steady-State Convergence Test Minimum | Steady-State Convergence Test Maximum | Distribution |
|---|---|---|---|---|---|---|---|
| $S_S$ | Soluble organic substrate | mg-COD·L$^{-1}$ | 5 | 5 | 2.5 | 7.5 | Uniform |
| $S_I$ | Soluble inert organic matter | mg-COD·L$^{-1}$ | 0 | 0 | - | - | - |
| $X_I$ | Particulate inert organic matter | mg-COD·L$^{-1}$ | 1,000 | - | 500 | 1,500 | Uniform |
| $X_S$ | Particulate organic substrate | mg-COD·L$^{-1}$ | 100 | - | 50 | 150 | Uniform |
| $X_{BH}$ | Active heterotrophic biomass | mg-COD·L$^{-1}$ | 500 | - | 250 | 750 | Uniform |
| $X_{BA}$ | Active autotrophic biomass | mg-COD·L$^{-1}$ | 100 | - | 50 | 150 | Uniform |
| $X_P$ | Particulate product from biomass decay | mg-COD·L$^{-1}$ | 100 | - | 50 | 150 | Uniform |
| $S_O$ | Dissolved oxygen | mg-O$_2$·L$^{-1}$ | 2 | 2 | 1 | 3 | Uniform |
| $S_{NO}$ | Nitrate and nitrite nitrogen | mg-N·L$^{-1}$ | 20 | 20 | 10 | 30 | Uniform |
| $S_{NH}$ | Ammonium nitrogen | mg-N·L$^{-1}$ | 2 | 2 | 1 | 3 | Uniform |
| $S_{ND}$ | Soluble biodegradable organic nitrogen | mg-N·L$^{-1}$ | 1 | 1 | 0.5 | 1.5 | Uniform |
| $X_{ND}$ | Particulate biodegradable organic N | mg-N·L$^{-1}$ | 1 | - | 0.5 | 1.5 | Uniform |
| $S_{ALK}$ | Alkalinity, assumed to be bicarbonate | mmol·L$^{-1}$ | 7 | 7 | 3.5 | 10.5 | Uniform |
| $TSS_1$ | Total suspended solids in layer 1 (top) | mg·L$^{-1}$ | - | 10 | - | - | - |
| $TSS_2$ | Total suspended solids in layer 2 | mg·L$^{-1}$ | - | 20 | - | - | - |
| $TSS_3$ | Total suspended solids in layer 3 | mg·L$^{-1}$ | - | 40 | - | - | - |
| $TSS_4$ | Total suspended solids in layer 4 | mg·L$^{-1}$ | - | 70 | - | - | - |
| $TSS_5$ | Total suspended solids in layer 5 | mg·L$^{-1}$ | - | 200 | - | - | - |
| $TSS_6$ | Total suspended solids in layer 6 | mg·L$^{-1}$ | - | 300 | - | - | - |
| $TSS_7$ | Total suspended solids in layer 7 | mg·L$^{-1}$ | - | 350 | - | - | - |
| $TSS_8$ | Total suspended solids in layer 8 | mg·L$^{-1}$ | - | 350 | - | - | - |
| $TSS_9$ | Total suspended solids in layer 9 | mg·L$^{-1}$ | - | 2000 | - | - | - |
| $TSS_{10}$ | Total suspended solids in layer 10 (bottom) | mg·L$^{-1}$ | - | 4000 | - | - | - |

## S3. Additional Uncertainty Analysis Results of BSM1

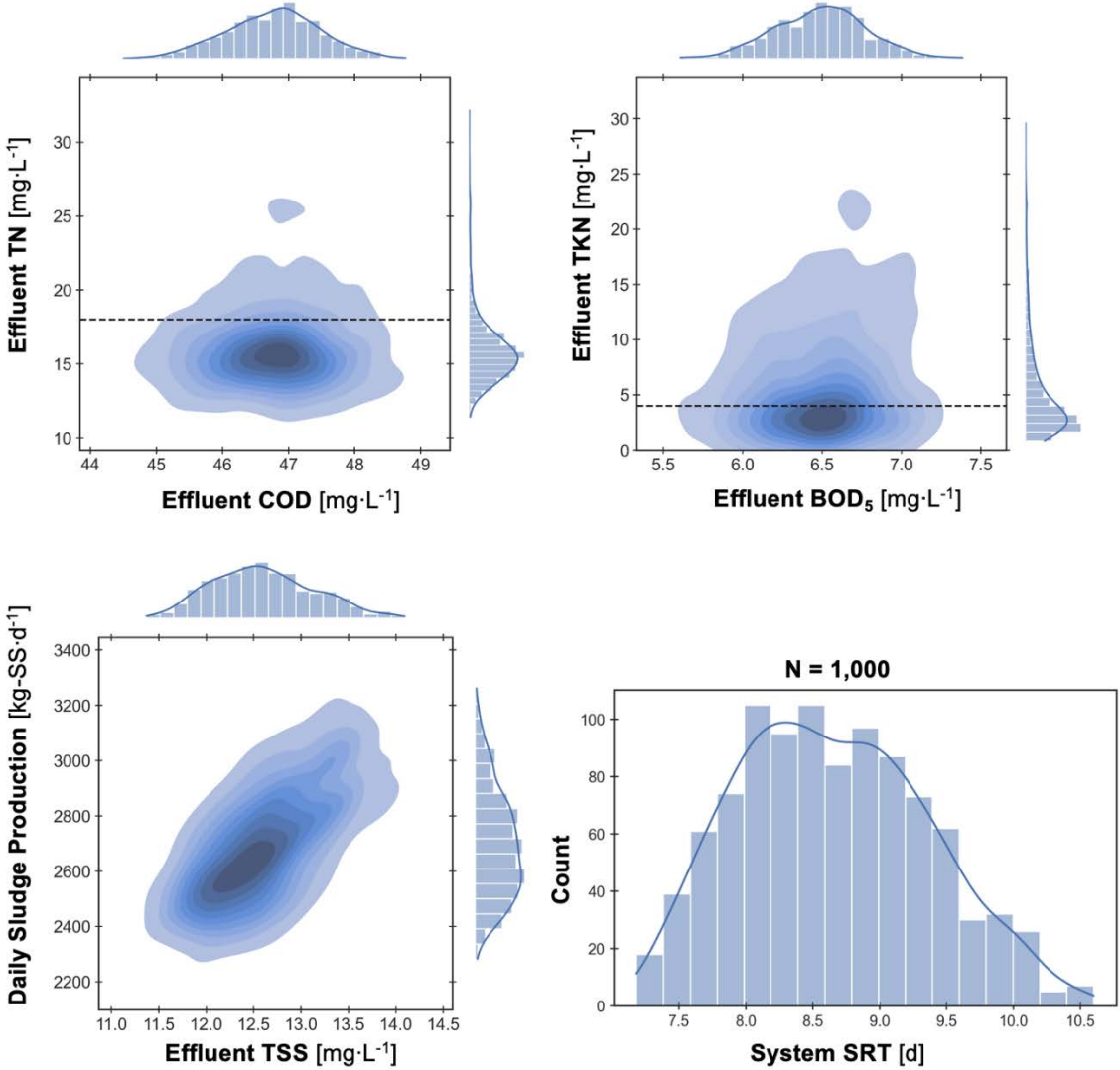

**Figure S3.** Kernel density plots of BSM1 system performance metrics at steady state. Black dashed lines indicate the assumed discharge limits of corresponding composite variables. The plots were created with the *stats* module in QSDsan.

## S4. Sensitivity Analysis (Monte Carlo Filtering) Results of BSM1

**Table S4**. Monte Carlo filtering results generated with simulation data from uncertainty analysis. A $p$ value less than 0.05 means the sample distributions of a variable significantly differ between the group with the metric values above the threshold (i.e., discharge limit) and the group below the threshold. The $D$ value indicates the "distance" between the variable's two sample distributions (*** $p < 0.001$; ** $p < 0.01$; * $p < 0.05$).

| Variable | Metric | | | |
|---|---|---|---|---|
| | Effluent TN | | Effluent TKN | |
| | $D$ | $p$ | $D$ | $p$ |
| ASM1 parameters | | | | |
| $Y_H$ | 0.077 | 0.568 | 0.042 | 0.746 |
| $Y_A$ | 0.086 | 0.422 | 0.058 | 0.358 |
| $f_{Pobs}$ | 0.073 | 0.622 | 0.102* | 0.011 |
| $i_{XB}$ | 0.062 | 0.806 | 0.050 | 0.545 |
| $i_{XP}$ | 0.140* | 0.035 | 0.055 | 0.420 |
| $f_{SS,COD}$ | 0.056 | 0.896 | 0.034 | 0.924 |
| $\mu_H$ | 0.113 | 0.140 | 0.080 | 0.081 |
| $K_S$ | 0.070 | 0.683 | 0.048 | 0.611 |
| $K_{OH}$ | 0.246*** | 0.000 | 0.090* | 0.034 |
| $K_{NO}$ | 0.117 | 0.117 | 0.083 | 0.063 |
| $b_H$ | 0.126 | 0.074 | 0.051 | 0.517 |
| $\mu_A$ | 0.173** | 0.004 | 0.145*** | 0.000 |
| $K_{NH}$ | 0.146* | 0.025 | 0.161*** | 0.000 |
| $K_{OA}$ | 0.168** | 0.006 | 0.114** | 0.003 |
| $b_A$ | 0.125 | 0.080 | 0.110** | 0.005 |
| $\eta_g$ | 0.142* | 0.032 | 0.081 | 0.073 |
| $k_a$ | 0.135* | 0.045 | 0.082 | 0.070 |
| $k_h$ | 0.155* | 0.015 | 0.058 | 0.358 |
| $K_X$ | 0.064 | 0.778 | 0.041 | 0.793 |
| $\eta_h$ | 0.081 | 0.492 | 0.057 | 0.380 |
| Decision variables | | | | |
| $V_a$ | 0.092 | 0.335 | 0.054 | 0.443 |
| $V_o$ | 0.251*** | 0.000 | 0.229*** | 0.000 |
| $K_L a_1$ | 0.512*** | 0.000 | 0.503*** | 0.000 |
| $K_L a_2$ | 0.066 | 0.752 | 0.085 | 0.054 |
| $Q_{RAS}$ | 0.108 | 0.178 | 0.167*** | 0.000 |
| $Q_{WAS}$ | 0.130 | 0.063 | 0.185*** | 0.000 |
| $Q_{intr}$ | 0.068 | 0.714 | 0.055 | 0.437 |
| Contextual parameters | | | | |
| $DO_{sat}$ | 0.263*** | 0.000 | 0.269*** | 0.000 |